\input{aipcheck}

\documentclass[
    ,final            
  ]
  {aipproc}

\layoutstyle{6x9}

\def\nin{\noindent}
\def\beq{\begin{equation}}
\def\eeq{\end{equation}}
\def\bea{\begin{eqnarray}}
\def\eea{\end{eqnarray}}

\begin{document}

\begin{flushright}
UGFT-279/10\\
CAFPE-149/10
\end{flushright}


\title{Hunting resonance poles with Rational Approximants}



 \author{Pere Masjuan}{
  address={Departamento de F\'{\i}sica Te\'{o}rica y del Cosmos, Universidad de Granada\\
  Campus de Fuentenueva, E-18071 Granada, Spain.\\
  masjuan@ugr.es}}

\begin{abstract}
\nin

Based on the mathematically well defined Pad\'{e} Theory, a theoretically safe new procedure for the extraction of the pole mass and width of resonances is proposed. In particular, thanks to the Montessus de Ballore's theorem we are able to unfold the Second Riemann sheet of an amplitude to search the position of the resonant pole in the complex plane. The method is systematic and provides a model-independent treatment of the prediction and the corresponding errors of the approximation.

This letter partially covers the material presented by the author at the $15^{th}$ International QCD Conference: QCD 10 (25th anniversary), Montpellier, France, 28 Jun - 3 Jul 2010 and at the Quark Confinement
and the Hadron Spectrum IX, 30 August - 3 September 2010, Madrid, Spain.
\end{abstract}

\maketitle


\section{Introduction}

\nin

The non-perturbative regime of QCD is characterized by the presence of physical resonances, complex poles of the amplitude in the transferred energy at higher (instead of the physical one) complex Riemann sheets. From the experimental point of view, one can obtain information about the spectral function of the amplitude through the Minkowsky region ($q^2>0$) and also about its low energy region through the experimental data on the Euclidean region ($q^2<0$). In reference \cite{MJJP-VFF}, the particular case of the $\pi\pi$ vector form factor (VFF) was analyzed with the main purpose of studying its low energy behavior using the available Euclidean data. In particular, the first and the second derivatives of the VFF were determined at $q^2=0$ with Pad\'{e} Approximants (PA) centered at the origin trough a fit procedure to that data \cite{MJJP-VFF}. In such a way, the vector quadratic radius $\langle r^2\rangle_V^{\pi}$ and the curvature $c_V^{\pi}$ were extracted from the fit and, as a consequence, a value for the low-energy constant $L_9=(6.84\pm0.07)\cdot 10^{-3}$ was obtained, \cite{MJJP-VFF,PhDMasjuan}.

Despite the nice convergence and the systematical treatment of the errors, this procedure does not allow us to obtain properties of the amplitude above the threshold, such as in the case of the $\pi\pi$ vector form factor, the $\rho$-meson pole position. The reason is simple: the convergence of a sequence of Pad\'{e} Approximants centered at the origin of energies ($q^2=0$) is limited by the presence of the $\pi-\pi$ production brunch cut, \cite{PhDMasjuan}. The PA sequence converge everywhere except on the cut. Still, the mathematical Pad\'{e} Theory allow us to produce a model independent determination of resonance poles when certain conditions are fulfilled. The most important one is to center our Pad\'{e} approximant sequence above the branch-cut singularity (beyond the first production threshold) instead of at origin of energies ($q^2=0$). This small modification also provides the opportunity to use Minkowskian data in our study instead of the Euclidean one. The relevance of this model-independent method to extract resonance poles is clear since does not depend on a particular lagrangian realization or modelization on how to extrapolate from the data on the real energy axis into the complex plane.

Although we apply this method in the particular case of a physical amplitude to extract the position of a resonance pole, is clear that it can be applied in a broader number of cases since only relies on a mathematical theory and not on a particular physical situation. We illustrate that method using an example where the properties above appear naturally.

Imagine a function $F(x)$ analytic in a disk $B_{\delta}(x_0)$. Then, the Taylor expansion $F(x)=\sum_{n=0}^{N} a_n (x-x_0)^n$ converges to $F(x)$ in $B_{\delta}(x_0)$ for $N\rightarrow \infty$, with derivatives $a_n=F^{(n)}(x_0)/n!$. In that situation, one usually uses experimental data to extract the derivatives of $F(x)$ using polynomial fits at higher and higher order $N$. Since the experimental data have errors, one normally finds that polynomials with order higher than a certain $N^*$ do not produce new information, with the new coefficients of order $N^*+1$ been compatible with zero. In that situation, one stops the fit procedure to order $N^*$ and believe that is the best one.
The scenario changes, however, when the function $F(x)$ is not analytic anymore, for example when has inside the disk $B_{\delta}(x_0)$ a single pole at $x=x_p$. In this case, the Taylor series does not converge any more, and we need a different procedure to extract information about the function.

The Montessus de Ballore's theorem states that the sequence of one-pole Pad\'{e} Approximants $P_1^N$ around $x_0$,

\begin{equation}\label{PAeq}
P_1^N(x,x_0)=\sum_{k=0}^{N-1}a_k(x-x_0)^k+\frac{a_N(x-x_0)^N}{1-\frac{a_{N+1}}{a_N}(x-x_0)}\, ,
\end{equation}

converges to $F(x)$ in any compact subset of the disk excluding the pole $x_p$, i.e,

\begin{equation}\label{th}
\lim_{N\rightarrow \infty} P_1^N (x,x_0)\, =\, F(x)\, .
\end{equation}

As an extra consequence of this theorem, one finds that the Pad\'{e} Approximant pole $x_{PA}=x_0+\frac{a_N}{a_{N+1}}$ converges to $x_p$ for $N\rightarrow \infty$ as it must be as stated by Eq. (\ref{th}). Since experiments provide us with values of $F_j$ at different $x_j$ instead of the derivatives of our function, we can use the rational functions $P_1^N$ as fitting functions to the data in a similar way as in Ref.~\cite{MJJP-VFF}. In this way, as $N$ grows $P_1^N$ gives us an estimation of the series of derivatives and the $x_p$ pole position.

Usually, as we have already said, Pad\'{e} Approximants are constructed around the low-energy point $x_0=0$ where $x$ is the total energy squared. For a physical amplitude, the function $F(x)$ (without a right-hand cut) is analytic from $x=-\infty$ up to the first production threshold $x_{th}$ and within the disk $B_{x_{th}}(0)$. In the $\pi \pi$ vector form factor case, the threshold is found to be at $x_{th}=4m_{\pi}^2$, where $m_{\pi}$ is the mass of the pion. Experiments then may provide with data $F^{exp}(x)$ at $x<0$ and use them to extract the derivatives of the vector form factor at the origin \cite{MJJP-VFF}. One may also have Minkowskian $F^{exp}(x+i0^+)$ data at $x>x_{th}$ which cannot be used to extract form factor properties with Pad\'{e} Approximants centered at $x_0=0$ due to the presence of the essential singularity at $x=x_{th}$. However, one can still use Pad\'{e} Approximants in a safe way by using the Montessus' theorem and center the approximants at $x_0+i0^+$ over the brunch cut between the first and the second production threshold, i.e., between $x_{th}<x_0<\tilde{x}_{th}$. In the $\pi\pi$ vector form factor that would correspond to the range between pion production threshold and the kaon one, $x_{th}=4m_{\pi}^2$ and $\tilde{x}_{th}=4m_K^2$ (assuming small multipion channels). In such a way we unfold the Second Riemann sheet due to the analytical extension of the function $F(x)$ from the first Riemann sheet at $x+i0^+$ into the second one.

In the case of resonant amplitudes, a single pole appears in the neighborhood of the real $x$ axis in the second Riemann sheet which can be related to a hadronic state, a resonance. Once we have unfold the Second Riemann sheet by locating our approximants over the brunch cut, the application of the Montessus' theorem in the disk of convergence (which is the region defined between the thresholds) is straightforward and allows us to locate the position of the resonant pole if lies inside that region. If that is the case, our $P_1^N$ Pad\'{e} approximant sequence determine systematically its position in a model independent way.

In the next section we present the details of this procedure by a particular example and in the third section we apply all this technology to the real case of the $\pi\pi$ vector form factor ALEPH data to extract the mass and the width of the $\rho$-meson.

\section{Testing the method with a model}
\nin

To illustrate the possibilities of our method we consider a $\rho$-like model of the $\pi\pi$ vector form factor, with a single pole in the second Riemann sheet at $q_p^2=(0.77-i0.15/2)^2$GeV$^2$ and a logarithmic branch cut (starting at $q^2=0$ for simplicity). The model:

\begin{equation}\label{rho-model}
F(q^2)=\frac{M^2}{M^2-q^2+\frac{1}{\pi}\frac{\Gamma q^2}{M}\ln \frac{-q^2}{M^2}}\, ,
\end{equation}

has two parameters, $M$ and $\Gamma$, tuned to produce the pole exactly at $q^2=q_p^2$.

There are two different ways to explore the method with this model. The first one consists on using the derivatives of the model to construct a $P_1^N$ Pad\'{e} Approximants sequence and extract from them the convergence sequence for the position of the pole. We call this method the \textit{genuine method}. The second method consists on simulating a physical situation generating a series of data points with zero error, which would represent an ideal experimental situation where all the uncertainty would be theoretical. We fit then that set of data for the modulus and the phase-shift of $F(q^2)$ with the corresponding modulus and phase-shift of each Pad\'{e} Approximant and we extract the complex parameter $a_n$ for each $P_1^N$. The Pad\'{e} Approximant pole $q_{fit}^2=(M_{fit}-i\Gamma_{fit}/2)^2$ is found to converge very quickly to the real $q^2_{p}=(M_{p}-i\Gamma_{p}/2)^2$ of the model as $N\rightarrow \infty$.

In the first case, using the genuine method, one first expand the function $F(q^2)$ at $q^2=q_0^2$, $F(q^2)=a_0(q^2-q_0^2)^0+a_1(q^2-q_0^2)^1+{\cal O}((q^2-q_0^2)^2)$, and with this expansion one construct the simplest Pad\'{e} Approximant $P_1^0(q^2;q_0^2)=\frac{a_0}{1-\frac{a_1}{a_0}(q^2-q_0^2)}$. Then, one looks for the pole of this $P_1^0$, $q_{P_1^0}^2=q_0^2+\frac{a_0}{a_1}$, and compares it with the pole position of the function $F(q^2)$. In a second step, one expands $F(q^2)$ at $q^2=q_0^2$ up to one order more, i.e, $F(q^2)=a_0(q^2-q_0^2)^0+a_1(q^2-q_0^2)^1+a_2(q^2-q_0^2)^2+ {\cal O}((q^2-q_0^2)^3)$, and constructs the approximant $P_1^1(q^2)=\frac{a_0+\frac{(a_1^2-a_0 a_2)(q^2-q_0^2)}{a_1}}{1-\frac{a_2}{a_1}(q^2-q_0^2)}$ to extract its pole $q_{P_1^1}^2=q_0^2+\frac{a_1}{a_2}$ and so on. To be able to appreciate the relevance of our approximation, we define the distance $dist$ between the predicted pole and the real pole as

\begin{equation}
dist=\sqrt{(M_{PA}-M_{pole})^2+(\Gamma_{PA}-\Gamma_{pole})^2}\, .
\end{equation}

\begin{figure}
\includegraphics[width=7cm]{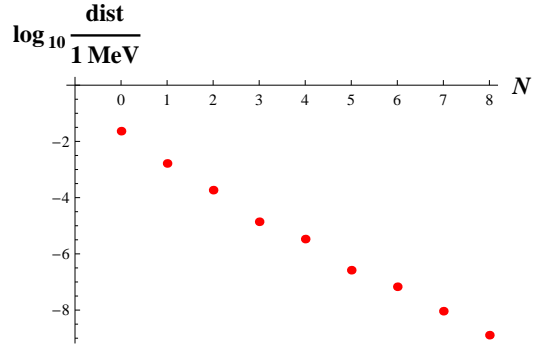}\\
  \caption{Rate of convergence of the sequence of \textit{genuine} $P_1^N$ Pad\'{e} Approximants corresponding to the first case studied for a $\rho$-like model (see the text for details).}\label{RC1}
\end{figure}

\begin{figure}
  \includegraphics[width=7cm]{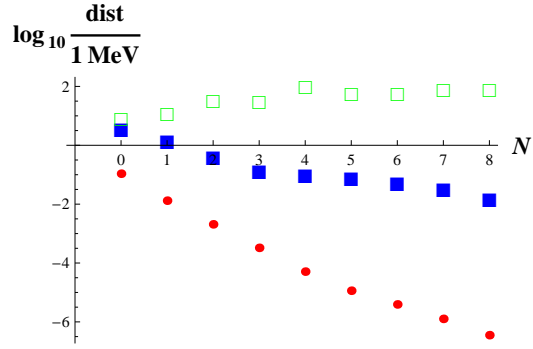}\\
  \caption{Rate of convergence of the $P_1^N$ Pad\'{e} approximants sequence corresponding to the $\rho$-like model (red-circles), the $\sigma$-like model (blue-filled squares) and the \textit{ultra-fat}-like model (green-empty squares).}\label{RC3}
\end{figure}

This parameter $dist$ helps us to see the rate of convergence of our sequence of approximants as is shown in Fig. \ref{RC1} where N is the order of our approximation. In particular, the first prediction, using the $P_1^0$, has an error below $0.1$MeV, the second one below $0.01$ MeV, and so on. The reader should take into account that the plot is in logarithm scale. At that point, a word of caution is needed. To succeed on the prediction of the position of a resonant pole using the Montessus' theorem a crucial condition must be fulfilled. The resonant pole we are looking for must lie within the disk of applicability of the theorem, in our case, in the disk limited by both production thresholds. In that sense, the prediction of the position of the pole in a $\rho$-like model converge very fast, for a $\sigma$-like model (which has a fatter resonance than a $\rho$-like model), the convergence is slower and for a ultra-fat-like model there are no convergence at all. This pattern of convergence is schematize in Fig. \ref{RC3} where red dots represent a $\rho$-like model with a resonance at $q^2_p=(0.77-i0.15/2)^2GeV^2$, the blue squares a $\sigma$-like model with a resonance at $q^2_p=(0.48-i0.53/2)^2GeV^2$ and the empty-green squares an ultra-fat model with a resonance at $q^2_p=(2.10-i1.05/2)^2GeV^2$. All in all, we can conclude that the difference between the prediction of the $q_p^2$ pole using the $P_1^N$ and using the $P_1^{N+1}$ gives us an estimation of the systematic error of our method.

\begin{figure}
  \includegraphics[width=7cm]{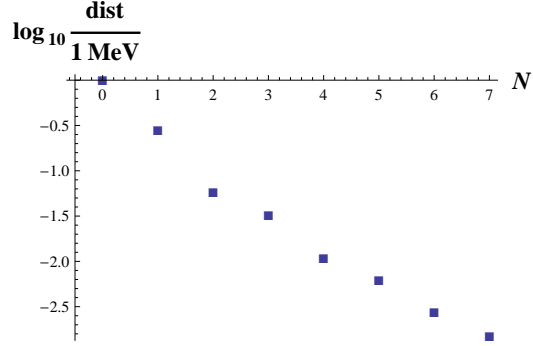}\\
  \caption{Rate of convergence of the fitted sequence $P_1^N$ Pad\'{e} Approximants corresponding to a $\rho$-like model considered in the second case of application of our mehtod (see text for details).}\label{RC2}
\end{figure}

As a second case of application of the Montessus' theorem, we generate a series of zero error data points for both modulus and phase-shift of our $\rho$-like model, Eq.~(\ref{rho-model}). We then construct a generic $P_1^N$ Pad\'{e} Approximant sequence which have several unknown parameters $a_n$. To predict the position of the resonant pole we need to know the values of these $a_n$ parameters. We fit then the modulus of our $P_1^N$ to the modulus of our $\rho$-like model and the phase-shift of our $P_1^N$ to the phase-shift of our $\rho$-like model. Again, when the $a_n$ parameters are known, we extract the position of the PA pole and we compare it with the real position through the distance $dist$. The rate of convergence of our new fitted sequence is shown in Fig. \ref{RC2} and gives us again an estimation of the systematic error of our method in this second case. As expected, this convergence is slower compare with the previous genuine case.

\section{Application of the method in a physical case}

We would like to use now our method to analyze the final compilation of ALEPH $\pi\pi$ vector form factor data for the squared modulus $|F_{\pi\pi}(q^2)|^2$, \cite{ALEPH}, and the $I=J=1$ $\pi\pi$ scattering phase-shift $\delta_{\pi\pi}$, identical to the $\pi\pi$ vector form factor phase-shift in the elastic region $4m_{\pi}^2<q^2<4m_K^2$ (if multipion channels are neglected), which conforms the range of applicability of our $P_1^N$ Pad\'{e} Approximant sequence (more details can be found in Ref. \cite{ProcJJ}). For $N\geq3$ the fit $\chi^2$ already lies within the 68$\%$ confidence level (CL) and becomes statistically acceptable.
At this point one needs to reach the typical fitting compromise. On one hand the experimental errors have an statistical origin and the contribution of this error increase as one considers higher order Pad\'{e} Approximants $P_1^N$, with larger number of parameters. On the other hand, the systematic theoretical error decreases as $N$ increases and the PA converges to the actual VFF. In the present work we have taken $N=6$ as our best estimate as the new parameters of Pad\'{e} Approximants with $N\geq7$ turn out to be all compatible with zero, introducing no more information with respect to the previous $P_1^6$. Furthermore, the model studied before shows that the theoretical errors for mass and width results are smaller than $10^{-2}-10^{-3}$MeV for $N\geq6$ (see Fig.~\ref{RC2}), being negligible compared to the ${\cal O}$(1 MeV) experimental errors. This yields to the determination:

\begin{equation}\label{result}
\centering
M_{\rho}=763.7\pm1.2 \mathrm{MeV}, \quad \Gamma_{\rho}=144\pm3 \mathrm{MeV}\, ,
\end{equation}

which is found to be in good agreement with previous determinations shown in Table \ref{tab:results} using more elaborated and complex procedures and with similar size of uncertainties .

{\footnotesize
\begin{table}
\centering
\begin{tabular}{|c|c|c|}
  \hline
  Ref. & $M_{\rho}(\mathrm{MeV})$ & $\Gamma_{\rho}(\mathrm{MeV})$\\
  \hline
  \cite{Ananthanarayan}  & $762.5\pm2  $&$ 142\pm7$ \\
  \cite{Pelaez}                             & $754  \pm18  $&$ 148\pm20$ \\
  \cite{Zhou}            & $763.0\pm0.2  $&$ 139.0\pm0.5$ \\
  \cite{PichCillero}        & $764.1\pm2.7^{+4.0}_{-2.5}  $&$ 148.2\pm1.9^{+1.7}_{-5.9} $ \\
  \hline
  this work                       & $763.7\pm1.2  $&$ 144\pm 3 $\\
  \hline
\end{tabular}
\caption{Comparison of different results for the determination of the mass and width of the $\rho$-meson.}\label{tab:results}
\end{table}
}

\section{Outlook}

A next step of this line of analysis would consist on incorporating another resonance in the problem where one should make use of the Montessus' theorem with a $P_2^N$ Pad\'{e} Approximant sequence instead of the $P_1^N$ one (or, more general, $P_L^N$ when a fixed number $L$ of resonances are include in the problem). However, one should take into account that we are not demanding the function to be meromorphic. In that case, it is better to use the Pomerenke's theorem as was shown in \cite{Masjuan}, instead of the method developed here. On the other hand, since we are dealing with functions that present a branch cut and pole(s), one can make use of the so-called Multivalued Approximants, or Non-Rational Approximants, which generalize the usual Pad\'{e} Approximants. Firstly considered by C. Pad\'{e} himself, did not receive much attention until the works of Shafer \cite{QA} and Short \cite{Short}. The simplest of these approximants are the so-called Quadratic Approximants (QA), defined by

\begin{equation}
F^2(x) \sum_{n=0}^{l} \gamma_n x^n + F(x) \sum_{n=0}^{k} \beta_n x^n + \sum_{n=0}^{j} \alpha_n x^n = {\cal O}(x^{j+k+l+2})\, .
\end{equation}

\begin{figure}
  \includegraphics[width=7cm]{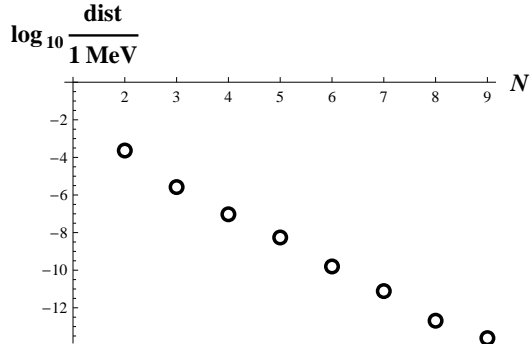}\\
  \caption{Rate of convergence of the $QA$ Quadratic Approximant sequence corresponding to the $\rho$-like model {\protect Eq.~\ref{rho-model}}.}\label{RC4}
\end{figure}

However, that can be extended to a given $k$ order just following the definition:

\begin{equation}
\sum_{n=0}^{N} F^k(x) \sum_{n=0}^{m_k} A_{k,n} x^n = {\cal O}(x^{\sum_{m_k}+N}) \, .
\end{equation}

where $k=2$ would correspond to the QA, $k=3$ to a Cubic Approximant and so on \cite{Short}. In our case, the order $N$ appearing in Fig.~\ref{RC4} is defined as $N=l+k+j$ with $N\geq2$ and $l\geq1$ in order the QA to has a brunch cut.

Such approximants possesses a build-in cut structure, they seem then ideally suited for problems related to multivalued functions such as our case. A preliminary study of the $\rho$-like model of Eq.~(\ref{rho-model}) with this kind of Quadratic Approximants centered again at $q_0^2$ turns out to be very promising considering the their rate of convergence shown in Fig.~\ref{RC4} where we have selected the best approximant for each $N$ among the $N+(N-1)+(N-2)+\cdots +1$ possibilities.

Finally, in several studies of S-matrix theory has been common to use, instead of the variable $q^2$, a variable $k$ related with the previous one through the relation $k=\sqrt{q^2-4m^2}$ (where $m$ is the produced particle mass). In the $k$-plane, the resonances are closer to the real axis than in the $q^2$-plane, so the extension of our method in this new variable suggests better pole position determination for those resonances with large width (such as the sigma meson).

\section{Conclusions}
\nin

We have develop a model-independent method for extracting resonance poles from physical amplitudes. The method is based on the well defined mathematical theory of Pad\'{e} Approximants and makes use of the Montessus' theorem to systematize the algorithm of extracting the desired resonance pole. However, our method has a larger application since does not rely on a particular lagrangian or in a modelization on how to extrapolate from the data on the real energy axis into the complex plane (such as other methods discussed in Ref. \cite{MSCV}). In such a way, the analysis of other available form factors and phase-shifts will be presented elsewhere.

In the particular case presented here we have analyzed the experimental Minkowskian $\pi\pi$-VFF and $\pi\pi$-scattering data by a $P_1^N$ Pad\'{e} Approximant sequence centered between the first and the second production thresholds (in such a way that we can unfold the second Riemann sheet of our amplitude). We have obtained a prediction of the $\rho$-meson pole position with compatible accuracy compared to other determinations of the same quantity (see Tab. \ref{tab:results}) but with a simpler and systematic method.

\begin{theacknowledgments}

This work has been performed in collaboration with Juan Jos\'{e} Sanz Cillero and has been supported by EU contract MRTN-CT-2006-035482 (FLAVIAnet), by MICINN, Spain (FPA2006-05294), the Spanish Consolider-Ingenio 2010 Programme CPAN (CSD2007-00042) and by Junta de Andaluc\'{\i}a (Grants P07-FQM 03048 and P08-FQM 101).
\end{theacknowledgments}



\bibliographystyle{aipproc}   




\end{document}